\documentstyle[aps,twocolumn]{revtex}

\begin{document}
\draft

\title{Magnetic-field dependence of the localization length in {Anderson}
insulators}

\author{Igor V. Lerner}
\address{School of Physics and Space Research, University of Birmingham,
Edgbaston, Birmingham~B15~2TT, United Kingdom}
\author{Yoseph Imry}
\address{Weizmann Institute of Science, Department of Condensed Matter Physics,
IL-76100 Rehovot, Israel\\ \medskip
Submitted to Europhysics Letters, July 30, 1994
%
\\ \medskip}\author{\small\parbox{14.5cm}{\small
 Using the conventional scaling approach as well as the renormalization group
 analysis in $d=2+\epsilon$ dimensions, we calculate the localization length
 $\xi(B)$ in the presence of a magnetic field $B$.
 For the quasi 1D case the results are consistent with a universal increase
 of $\xi(B)$ by a numerical factor when the magnetic field is
 in the range $\ell\ll{\ell_{\!{_H}}}\alt\xi(0)$,
 $\ell$ is the mean free path, ${\ell_{\!{_H}}}$ is the magnetic length
 $\sqrt{\hbar c/eB}$.
 However, for $d\ge 2$ where the magnetic field does cause
 delocalization there is no
 universal relation between $\xi(B)$ and $\xi(0)$. The effect of
 spin-orbit interaction is briefly considered as well.
}}\address{}\maketitle

%

%
\bibliographystyle{simpl1}

 \narrowtext
 Different universality classes,  introduced by Dyson \cite{Dsn}, became
 the framework for considering the role of symmetry breaking (due to the
 presence of a magnetic field, or spin-orbit scattering)
 in transport properties of disordered electron systems.
 This role has been well understood in the
 weak-localization regime, and in the vicinity of the  Anderson
 metal-insulator transition \cite{AALR,Weg:79,LR}.
 On the other hand, the
 classification in terms of the universality classes appears to be quite
 useful also for quasi-one-dimensional systems
 in the strongly localized regime.
 One of the intriguing predictions for this regime
 \nocite{PichSSD}\cite{EfL,Dor,SMMP,SMP}
 is the  existence of the universal
 relation, $\xi_\beta\!=\!\beta\times\xi_{\beta=1}$,
 between localization lengths in different universality classes characterized
 by the parameter  $\beta$.
 There is, however, some controversy:
 while the transfer-matrix approach based on the
 maximal entropy ansatz \cite{SMMP,SMP} predicts that  $\beta=1,\,2,$ or $ 4$
 for orthogonal, unitary,
 and symplectic ensembles respectively,
 in accordance with the Dyson's definition for the random-matrix theory
 \cite{RMT}, the exact calculation within the nonlinear $\sigma$ model
 \cite{EfL} shows that were spin-rotation symmetry  broken,
 $\beta$ would be equal $4$
 even if  time-reversal symmetry were broken as well, although such an ensemble
would behave as unitary ($\beta=2$) in relation to the level
statistics\cite{EfL}.  Nevertheless, the existence of
a universal relation for $\xi$  has been  reliably
 established for quasi-one-dimensional systems,
  including quantum-chaotic ones
 \cite{Smil}. However, the extension
of  the transfer-matrix approach to a higher dimensionality,
as in Ref.\ \cite{SMP}, is
not obviously  justifiable.

 In the present paper we analyze the influence of a
 magnetic field $B$ and spin-orbit scattering (at $B=0$)
 on the localization length
 $\xi$ applying  the standard one-parameter scaling and
 renormalization group (RG) approach \cite{LR}.
 The advantage of such a simple approach is  possibility to
 apply the very same scheme
 to any  dimensionality, $d$.
 We show that, while being in a {\it qualitative} agreement with the
 above-mentioned universality for $\xi$ in quasi-1D systems, the scaling
 predicts the absence of any universal relation for $d\ge 2$.

 The RG equation for the
 dimensionless conductance $g(L)$ is
\begin{eqnarray}
 \label{RG}
{d\ln g}/{d\ln L}
 =\beta(g).
 \end{eqnarray}
  Here $g(L)$ is defined
for  a $d$-dimensional system as the
conductance  in   units of  $\alpha_d (e^2/\pi\hbar)$
 with $\alpha_d\equiv2S_{d-1}/(2\pi)^d$, $S_d$ being the area of the
$d$-dimensional unit sphere.
  The $\beta$ function
  depends on the  time-reversal and spin-rotational symmetry of the problem.
Let us denote the $\beta$ functions
  at $B=0$, at large $B$ and in the presence of ``strong''
spin-orbit scattering (at $B=0$) respectively as $\beta_o$,
 $\beta_u$ and $\beta_s$. The subscripts $o$, $u$  and $s$ stand for
orthogonal,
  unitary and  symplectic
 symmetries, respectively.

 The question is  when is the symmetry effectively broken.
 The answer  depends on which length scale, $L$, we
 are in. We start at the microscopic scale $L\sim \ell$, with $\ell$ being
 the mean free path, and go up to the range $L\alt\xi$. At $d>2$,
 that includes both localized and extended phases.
 In the extended phase
 $\xi$ has the meaning of the correlation length which is indistinguishable
 in the critical regime, $L\alt\xi$, from the localization length in the
 insulating phase.  At $d\le2$, localization sets in with $L$ increasing
 up to and beyond $\xi$.

  In a  semiclassical approximation, typical electron trajectories have phases
 $({2\pi\phi/\phi_{_0}})$ (with $\phi_{_0}=hc/e$ being the flux quantum)
 which are negligible
 for small fluxes, when
$L\ll {\ell_{\!{_H}}}\equiv\sqrt{\hbar c/eB}$  and the problem is
 effectively that of $B=0$. On the other hand, once $BL^2\agt\phi_{_0}$, the
 phases are significant and the time-reversal symmetry is effectively broken.
 As long as localization has not set in, most of the relevant trajectories
 cover the whole area of the system, and increasing $B$ beyond
 ${\ell_{\!{_H}}}\sim L$ should not further  change the
 symmetry of the problem.
 Similar considerations
 may well be valid for general Feynman paths, beyond the semiclassical
 approximation\footnote{
 We do not consider here the range of larger magnetic fields
${\ell_{\!{_H}}}\alt\ell$,
  where this treatment breaks down as the
  classical trajectories bend and orbit shrinking starts. So we leave out of
  account interesting effects that may occur in the range of even higher
  fields, ${\ell_{\!{_H}}}^{ 2}\alt\ell/k_{_F}$, where one has to scale
$g_{xx}$ and
  $g_{xy}$ together\cite{DK1}\nocite{DK2}.
}. Thus we are led to the following assumption:
 {\it The correct $\beta(g)$ (for $g\agt 1, L\alt\xi$) can be approximated
 by $\beta_o$ for $L\alt{\ell_{\!{_H}}}$ and by $\beta_u$ for
$L\agt{\ell_{\!{_H}}}$}.

Similarly, we assume that for weak spin-orbit scattering at $B\!=\!0$,
 $\beta(g)$ is approximated by $\beta_o$ for
$L\alt\ell_{\text{so}}$, and by $\beta_s$ for
 $L\agt\ell_{\text{so}}$ ($\ell_{\text{so}}$ is the length  of
 spin-orbit scattering).

 The above assumption can be easily used in the RG procedure to estimate the
 change of $\xi$ as a function of $B$. Let us start at $d\le 2$ for {\it weak
 disorder}, $k_{_F}\ell\gg 1$, where localization still exists for large $L$.
 We start at a length scale $L_{_0}$, satisfying $\ell\alt L_{_0}\ll\xi$,
 where $g\gg 1$ and integrate the RG
 equation until getting to the regime where $g\sim 1$. This regime
corresponds to the scale
 $L\sim\xi$, and
 this is how $\xi$ is determined.

We approximate $\beta(g)$ by the
 weak-localization expres\-sions\cite{BHZJ,Weg:80a}:
 \begin{mathletters}
  \label{bb}
 \begin{eqnarray}
 \label{bo}
 \beta_o(g)&=&d-2-{1/ g}
 \,,\\[4pt]
 \label{bu}
  \beta_u(g)&=&d-2-
  {1/ 2g^2}\,,\\[4pt]
 \label{bs}
 \beta_s(g)&=&d-2+
 {1/ 2g}\,.
 \end{eqnarray}
 \end{mathletters}
The exact matching of
 $\beta(g)$ at $L\sim{\ell_{\!{_H}}}$ or
$L\sim\ell_{\text{so}}$ yields only unimportant corrections. We do not consider
the case where both time-invariance and spin-rotations symmetries are broken.

 We begin with a quasi $1D$ wire, having $N_{_\perp}$ conduction channels
 (i.e.\
 $N_{_\perp}$ transfer states below $E_F$).
 For the wire of a $2D$ cross section the classical Drude conductance
is given
 by
 \begin{eqnarray}
 \label{drude}
  g_{c\ell}(L)={\pi\over 3}  N_{_\perp}{\ell\over L}\,.
 \end{eqnarray}
 We start integration of Eq.\ (\ref{RG})
 from a small $L_{_0}\ll\xi$ (but $L_{_0}\gg\ell$).
For $B$ large
 enough, ${\ell_{\!{_H}}}\alt \xi_o$ ($\xi_o$ is the $B=0$
 localization length), we first integrate
 Eq.\  (\ref{RG}) in the orthogonal symmetry range, from $L_{_0}$ to
${\ell_{\!{_H}}}$,
 with the
 $\beta$-function $\beta_o$, Eq.\
 (\ref{bo}), to obtain (using
 $g_{c\ell}(L_{_0}\gg 1$)
 \begin{eqnarray}
 \label{1d0}
 g(L)+ 1=
 g_{c\ell}(L),\qquad L\alt{\ell_{\!{_H}}}\,.
 \end{eqnarray}
 Note that $L_{_0}$ has cancelled out. For $L$'s above ${\ell_{\!{_H}}}$,
 starting with
 $L={\ell_{\!{_H}}}$ we integrate Eq.\ (\ref{RG}) with the
 $\beta$-function $\beta_u$, Eq.\ (\ref{bu}),
 to find (using $g({\ell_{\!{_H}}})\gg 1$)
 \begin{eqnarray}
 \label{1du}
  g^2(L)+{1/2}=
g_{c\ell}^2(L),\qquad L\agt{\ell_{\!{_H}}}\,.
 \end{eqnarray}
 Note that ${\ell_{\!{_H}}}$ has cancelled out as well, this will be seen to be
 the source of the universal increase of $\xi$ with B in the whole regime
 considered in the quasi $1D$ case.

For a very weak magnetic field,
 ${\ell_{\!{_H}}}\gg\xi_o$, Eq.\ (\ref{1d0}) is valid
 in the whole range of $L$, up to $\xi_o$, and the behavior is
``orthogonal" throughout, still
 allowing small $B$-dependent corrections to $\xi$. At
 $L=\xi_o$, $g(\xi_o)=A$, $A$
 being a constant of order unity. Thus
 \begin{eqnarray}
 \label{xi1o}
  \xi_o= \frac{\pi}{3}{N_{_\perp}\ell\over A+1}\;,
 \end{eqnarray}
 in accordance with
 the well-known Thouless result \cite{DT:77}. In the purely
 unitary case, i.e.\ for a ``strong'' magnetic fields
 ${\ell_{\!{_H}}}\sim L_{_0} $, one has an {\it increase} of $\xi$:
 \begin{eqnarray}
 \label{xi1u}
 \xi_u=\frac{\pi}{3}{ N_{_\perp}\ell\over\sqrt{A^2+{1/2}}}\; .
 \end{eqnarray}
 A similar increase takes place in the
presence of   spin-orbit scattering.
Repeating the same procedure as above,
one finds that for the purely symplectic case
 \begin{eqnarray}
 \label{xi1s}
  \xi_s= \frac{\pi}{3}{N_{_\perp}\ell\over A-1/2}\;,
 \end{eqnarray}
 While the precise value of $\xi_u/\xi_o$ and $\xi_s/\xi_o$ cannot be
 determined by this simple approach,
 this increase is qualitatively due to $\beta(g)$ being {\it
 larger} in the unitary case, and even more so in the symplectic case
 so that one has to go to larger scales to get $g\sim 1$.  
 Since ${\ell_{\!{_H}}}$ dropped
 from Eq.\ (\ref{1du}),  the  increase from $\xi_0$ to $\xi_u$
 is {\it universal} (i.e.\ it does
 not depend on $B$) in the whole range $\ell\alt{\ell_{\!{_H}}}\alt\xi$.
Similarly,
the increase of $\xi$ in the symplectic case does not depend on the
 spin-orbit scattering strength for  $\ell\alt\ell_{\text{so}}\alt\xi$.
 Thus our
 scaling consideration, while not able to
 yield the constants, is consistent with the exact result \cite{EfL} that in a
 quasi $1D$ metallic wire there is an increase of $\xi$ in the
 appropriate, rather
wide, range of $B$.

   The physical reason for the increase of $\xi$ with $B$ is now very
 clear. It is due to the well-known delocalization for the weakly localized
 regime $L\ll\xi_o$, caused by the elimination of coherent time-reversed paths
 by the magnetic field $B$. The scaling procedure enables us to approximately
 carry this effect all the way up to $L\sim\xi$ and see by how much $\xi$ is
 changing.

 The same procedure which leads to semi-quantitatively correct results in
 quasi 1D and illuminates the physics of delocalization by the field,
 can be repeated for a $2D$ film.
 We start here at
 $L_{_0}\sim\ell$
 with a conductance $g_{_0}$ and obtain for $\xi_o\agt{\ell_{\!{_H}}}\agt\ell$
 \begin{eqnarray}
 \label{xi2}
 \xi={\ell_{\!{_H}}}\exp\!\left(g_{_H}^{2}\right)
 \; ,
 \end{eqnarray}
 where $g_{_H}\equiv g({\ell_{\!{_H}}})=g_{_0}- \ln ({\ell_{\!{_H}}}/\ell)$.
This ranges from the
 orthogonal value given by $g_{_0}\sim \ln({\xi_o/\ell})$, i.e.\
 $\xi_o\cong\ell e^{g_{_0}}$
 for ${\ell_{\!{_H}}}\agt\xi_o$, to the unitary one, $\xi_u\cong\ell
e^{g^2_{_0}}$
 for ${\ell_{\!{_H}}}\sim\ell$.
 Since $g_{_0}\gg 1$, $\xi_u$ is overwhelmingly larger than $\xi_o$.
 The ratio
 $\xi_u/\xi_o$ is obviously not universal -- it changes by many orders of
 magnitude in the range $\ell\alt{\ell_{\!{_H}}}\alt\xi_o$.

In the symplectic case, the $\beta_s$ function, Eq.\ (\ref{bs}),
is always positive in the weak localization regime.  So for $\xi_o<
\ell_{\text{so}}$, there is an ``antilocalization'' effect: the conductance is
decreasing with increasing the scale up to $L\sim \ell_{\text{so}}$. Then
the $\beta$ function in Eq.\ (\ref{RG}) changes sign from negative,
Eq.\ (\ref{bo}), to positive, Eq.\ (\ref{bs}),
and $g$ begins to increase. The antilocalization character of the $\beta$
 function might  change at $g\sim 1$
due to  further terms of the $1/g$ expansion
\cite{Weg:89}. Such a change, if any, is definitely beyond the scope
of the present treatment.

  We now consider the situation for $d>2$, where the Anderson
 transition exists at a certain $g=g_c$. Here we use the  RG
 equation (\ref{RG}) linearized near $g_c$ (for $g-g_c\ll 1$)
 with
 \begin{eqnarray}
 \label{lin}
  \beta(g)={1\over\nu}\,{g-g_c\over{g_c}},
 \end{eqnarray}
 where $\nu$ is the exponent of
 the diverging $\xi$ near the transition. Indeed, the solution to
 Eq.\ (\ref{RG})  with the linearized $\beta$-function (\ref{lin})
 for $g$ ($=g(L)$) at a general scale $L$ is written as:
 \begin{eqnarray}
 \label{sol1}
  g_c-g=(g_c-g_{_0})\left({L/ L_{_0}}\right)^{1/\nu}
 \; ,
 \end{eqnarray}
 where $g_{_0}=g(L_{_0})$. We rewrite this in the localized phase
 ($g_{_0}<g_c$) as
\begin{eqnarray}
 \label{sol}
 g=g_c \left[1-\left( {L}/{\xi}\right)^{1/\nu}\right],\qquad
 \xi\equiv L_{_0}\left({1-g_o/g_c}\right)^{ -\nu}\, .
 \end{eqnarray}
 From now on we take the above notation, with the constants $g_c$ and $\nu$,
 to describe the orthogonal case ($B=0$) with $\xi=\xi_o$. The unitary
 (``large $B$") case satisfies the same scaling theory but with the constants
 $g_c$ and $\nu$ replaced by
 ${g_{c{\scriptscriptstyle H}}}$
 and $\nu_{_H}$. In the $\epsilon$
 expansion for $d=2+\epsilon$, the transition at small $\epsilon$
 occurs in a range describable by perturbation theory, and we can use
 the weak-localization expressions (\ref{bb}) to obtain the following
 constants
 \begin{eqnarray}
 \label{CP}
  g_c= {1}/{\epsilon}, \;
 \nu={1}/{\epsilon};\;
{g_{c{\scriptscriptstyle H}}}=({1}/{2\epsilon})^{1/2},\;\qquad
{\nu_{\!{_H}}}={1}/{2\epsilon}\,.
 \end{eqnarray}
 We shall assume that
${g_{c{\scriptscriptstyle H}}}
<g_c$ also at $\epsilon$=1 and use our
 ansatz that the system behaves according to orthogonal or unitary symmetry
 for $L<{\ell_{\!{_H}}}$ or $L>{\ell_{\!{_H}}}$ respectively. We shall also
assume
 that the crossover is describable
 by the linearized $\beta(g)$. The latter assumption, too, is justified in the
 $\epsilon$-expansion. We do the considerations below for a general $d>2$.
 They are certainly valid for $\epsilon\ll 1$ and should be qualitatively
 correct at $d=3$. (Unfortunately,
such an approach is not feasible
for the spin-orbit case where $g_c$ is not large even for small $\epsilon$,
and the $\epsilon$-expansion could hardly provide with
useful insight into the
$3D$ case, besides the fact, though, that
the spin-orbit interaction still does
cause delocalization for $\ell\ll\ell_{\text{so}}\ll\xi$ in
the weakly localized phase near the transition.)

 Again, for $\ell\alt{\ell_{\!{_H}}}\alt\xi$, we use the orthogonal results,
 Eqs.\ (\ref{sol1}), (\ref{sol}), from $L_{_0}=\ell$ to ${\ell_{\!{_H}}}$:
 \begin{eqnarray}
 \label{gH}
 g({\ell_{\!{_H}}})=g_c-(g_c-g_{_0})\left({{\ell_{\!{_H}}}/\ell}\right)^
\frac{1}{\nu}\!
 =g_c \left[1-\left({{\ell_{\!{_H}}}/\xi_o}\right)^
\frac{1}{\nu}
\right]
 \end{eqnarray}
 At the scale $L\agt{\ell_{\!{_H}}} $,   we use the unitary results.
 When $g({\ell_{\!{_H}}})>
{g_{c{\scriptscriptstyle H}}}
 $, the decrease in $g$, Eq.\ (\ref{gH}),
 is changed by an increase as the unitary $\beta$-function (\ref{lin}) is
positive at this scale.
 Then,  the critical conductance as function of $B$,
 $g_c(B)$, is given by $g_{_0}$ for which $g({\ell_{\!{_H}}})=
  {g_{c{\scriptscriptstyle H}}}$.
Substituting into this equation $g({\ell_{\!{_H}}})$, Eq.\ (\ref{gH}), and
 solving for $g_{_0}=g_c(B)$ we find:
 \begin{eqnarray}
 \label{gch}
 g_c(B)=g_c-(g_c-{g_{c{\scriptscriptstyle H}}})
 \left({\ell/{\ell_{\!{_H}}}}\right)^\frac{1}{\nu}
 \end{eqnarray}
  To check this, we see that $g_c(B)=
 {g_{c{\scriptscriptstyle H}}}$
 for ${\ell_{\!{_H}}}=\ell$, and \mbox{$g_c(B)\rightarrow g_c$} for
  ${\ell_{\!{_H}}} \rightarrow\infty\; (B\rightarrow 0)$. The real
 divergence of $\xi$ near the transition occurs
 for a given $B$ when $g_c(B)\rightarrow g_{_0}$:
 \begin{eqnarray}
 \label{xiH}
 \xi_{_H}&=&{\ell_{\!{_H}}}\left(1-{g({\ell_{\!{_H}}})\over
 {g_{c{\scriptscriptstyle H}}} }\right)^{\!\!-\nu_{_H}} %
\nonumber\\&=& 
 {\ell_{\!{_H}}}
 \left({{\ell_{\!{_H}}}\over\ell}\right)^{\!\!-\nu_{_ H}/\nu}
 \left({g_c(B)-g_{_0}\over
 {g_{c{\scriptscriptstyle H}}}
   }\right)^{\!\!-\nu_{_H}}
 \end{eqnarray}
 At this scale, $g$ is given by Eq.\ (\ref{sol})
 with substituting
${g_{c{\scriptscriptstyle H}}},\,
 \xi_{_H}$, and $\nu_{_H}$ for respectively $g_c,\xi_o$, and $\nu$. Clearly,
there is nothing universal in the ratio of $\xi_H/\xi$. A similar conclusion
for $d=2,3$, both for the magnetic and spin-orbit case, has been made in Ref.\
\cite{Bouch1}\nocite{Bouch2}
based on the heuristic picture of localization of Ref.\ \cite{Bouch3}.

 The most
 important aspect of delocalization by a magnetic field in the range
 considered ($\ell\alt{\ell_{\!{_H}}}\alt\xi_o$) at $d>2$ is the shift of the
 transition towards delocalization. This is seen from the fact that for small
 $\epsilon$, ${g_{c{\scriptscriptstyle H}}}
 \ll g_c$ and it is expected that ${g_{c{\scriptscriptstyle H}}}<g_c$ also
 at $d$=3. For a given field B there are three regimes: for $g\le g_c(B)$,
 there is a localization also with the magnetic field but $\xi$ is larger
 (arbitrarily larger near $g_c(B)$); for $g_c(B)\le g\le g_c$ the magnetic
 field destroys localization completely; for $g>g_c$, there is no
 localization but the correlation length $\xi$ appears to be larger at $B=0$
 (especially just above $g_c$). The shift of the transition with field has
 been discussed in Refs.\ \cite{LKh,Sha:84}.
 Clearly, the effect of $B$ for $d\ge2$ is not universal
 (except, of course, that the asymptotic critical behavior
 is governed by the unitary universality class for any $B$).
 It is only for quasi 1D
 systems that $\xi$ increases by a universal factor in the whole range
 $\ell\alt{\ell_{\!{_H}}}\alt\xi_o$. It is interesting that the considerations
based
 just on the symmetry breaking within the transfer matrix approach
 \cite{SMMP} seem to be
 valid only for quasi 1D systems. They had, however, highlighted the very
 interesting and at the time almost unexpected possibility of
 substantial delocalization
 by the magnetic field.

The scaling cross-over considerations exemplified here can be applied to
various physical situations involving the phase-breaking length
$\ell_\phi$, alongside with  $\ell_{_{\!H}}$ and $\ell_{\text{so}}$ in
appropriate ranges of the localized phase. This  produces a variety of
interesting magnetoconductance phenomena in the localized phase.

{ \acknowledgments
 The authors thank  Y.\ Gefen, O.\ Entin-Wohl\-man,
 D.E.\ Khmel\-nitskii, Z.\ Ovadyahu, U.\ Sivan and U.\ Smilansky for
discussions
 on this and related topics. We are grateful to the Wissenschaftskolleg
in Berlin for kind hospitality extended to us at the initial stage of this
work.
This work was
supported by the Minerva Foundation, by the German-Israeli Foundation (GIF),
and by the EEC grant No.\ SSC-CT90-0020.
}


%

\end{document}